\documentclass[twocolumn,  
showpacs,  
preprintnumbers,  
aps,  
prd,  
a4paper,  
nofootinbib,  
tightenlines,  
floats, floatfix  
]{revtex4}

\usepackage{amsmath}
\usepackage{amsfonts}
\usepackage{graphicx}

\newcommand{\bea}{\begin{eqnarray}}
\newcommand{\ena}{\end{eqnarray}}

\newcommand{\R}{\cal{R}}
\renewcommand{\b}{\beta}
\renewcommand{\a}{\alpha}
\renewcommand{\c}{\gamma}
\renewcommand{\o}{\omega}

\begin{document}

\preprint{BI-TP 2009/15}

\title{Power spectra from an inflaton coupled to the Gauss-Bonnet term}

\bigskip

\author{Zong-Kuan Guo}
\email{guozk@physik.uni-bielefeld.de}
\affiliation{Fakult{\"a}t f{\"u}r Physik, Universit{\"a}t Bielefeld,
Postfach 100131, 33501 Bielefeld, Germany}

\author{ Dominik J. Schwarz}
\email{dschwarz@physik.uni-bielefeld.de}
\affiliation{Fakult{\"a}t f{\"u}r Physik, Universit{\"a}t Bielefeld,
Postfach 100131, 33501 Bielefeld, Germany}

\date{\today}

\begin{abstract}

We consider power-law inflation with a Gauss-Bonnet
correction inspired by string theory. We analyze the stability of
cosmological perturbations and obtain the allowed parameter space. We
find that for GB-dominated inflation ultra-violet instabilities of
either scalar or tensor perturbations show up on small scales.
The Gauss-Bonnet correction
with a positive (or negative) coupling
may lead to a reduction (or enhancement) of the tensor-to-scalar ratio
in the potential-dominated case.
We place tight constraints on the model parameters by making use of the 
WMAP 5-year data.
\end{abstract}


\maketitle

\section{Introduction}
Inflation in the early universe has become the standard model
for the generation of cosmological perturbations in the universe, 
the seeds for large-scale structure and temperature anisotropies of
the cosmic microwave background (CMB).
The simplest scenario of cosmological inflation is based on the presence
of a single, minimally-coupled scalar field with a slowly varying potential.
Quantum fluctuations of this inflaton field give rise to an almost scale-invariant 
and primordial power spectrum of isentropic perturbations 
(see Refs.~\cite{lyt99,bas05} for reviews).

String theory is often regarded as the leading candidate for unifying
gravity with the other fundamental forces and for a quantum theory of
gravity. It is known that there are correction terms of higher orders in the
curvature to the lowest order effective supergravity action coming from
superstrings, which may play a significant role in the early universe.
The simplest such correction is the Gauss-Bonnet (GB)
term in the low-energy effective action of the heterotic string~\cite{gos87}.
Such a term provides the possibility of avoiding the initial singularity of
the universe~\cite{ant93}.
In the presence of an exponential potential for the modulus field, 
nonsingular cosmological solutions were found which begin in an 
asymptotically flat region, undergo super-exponential inflation and 
end with a graceful exit to a phase with decreasing Hubble radius~\cite{tsu02}.

There are many works discussing accelerating cosmology with the GB
correction in four and higher dimensions~\cite{guo07a,sat07,noj05,bro05}.
Recently it has been shown that the GB term might give rise to violent 
instabilities of tensor perturbations~\cite{guo07}.  A model in which inflation is 
driven by the GB term and a higher-order kinetic energy term was studied. 
When the GB term dominates the dynamics of the background, tensor perturbations 
exhibit violent negative instabilities around a de Sitter background on small scales,
in spite of the fact that scale-invariant scalar perturbations can be
achieved~\cite{guo07}. Besides the kinetic and GB terms, a scalar potential arises naturally 
from supersymmetry breaking or other non-perturbative effects.
So far, the inflationary solutions and the resulting cosmological
perturbations have not been studied in detail when both the GB correction and
the scalar potential are present.

In this work we investigate single-field inflation with a non-minimal coupling of the 
inflaton to the GB term. We confront the predictions for the primordial power spectra 
for scalar and tensor modes with WMAP data. In doing so, we restrict our attention to 
power-law inflation, which is realized when both the potential and
the scalar-GB coupling take an exponential form. When the potential
is dominant, we find that it is possible to realize observationally
supported density perturbations. When the GB term is dominant,
either tensor or scalar perturbations exhibit negative instabilities
on small scales, which invalidate the assumption of linear perturbations. 

This paper is organised as follows. In section II we calculate the power spectra of scalar 
and tensor perturbations for an inflaton that is coupled to the GB term. We restrict our analysis 
to situations in which the mode equations are solved by Bessel functions. In section III we study
the case of power-law inflation in detail. Section IV is devoted to conclusions.

\section{Cosmological perturbations}
We consider the following action
\bea
S = \int d^4x\sqrt{g} \left[\frac12 R - \frac{\o}{2}(\nabla \phi)^2
 - V(\phi)-\frac12\xi(\phi) R_{\rm GB}^2\right],
\label{action}
\ena
where $\phi$ is a scalar field with a potential $V(\phi)$, $\o=\pm 1$,
$R$ denotes the Ricci scalar, and 
$R^{2}_{\rm GB} = R_{\mu\nu\rho\sigma} R^{\mu\nu\rho\sigma}
- 4 R_{\mu\nu} R^{\mu\nu} + R^2$ is the GB term.
We work in Planckian units, $\hbar = c = 8\pi G = 1$.
In a spatially flat Friedmann-Robertson-Walker (FRW) universe with scale
factor $a$, the background equations read
\bea
\label{beq1}
&& 6H^2 = \o \dot{\phi}^2 + 2V + 24\dot{\xi}H^3, \\
&& 2\dot{H} = -\o\dot{\phi}^2 + 4\ddot{\xi}H^2 + 4\dot{\xi}H (2\dot{H} -
H^2),
\label{beq2}
\ena
where a dot represents the time derivative and $H=\dot{a}/a$ denotes the expansion rate.
To compare the contributions from the potential and the GB term, we
use the ratio of the second to the third term on the right-hand
side of Eq.~(\ref{beq1}), $\lambda=V/12\dot{\xi}H^3$.
For a potential-dominated model $|\lambda|>1$, for a GB-dominated
model $|\lambda|<1$.

At linear order in perturbation theory, the Fourier modes of curvature
perturbations satisfy~\cite{car01}
\bea
v'' + \left(c_{\R}^2 k^2 - \frac{z''_{\R}}{z_{\R}}\right)v = 0,
\label{spm}
\ena
where a prime represents a derivative with respect to conformal time
$\tau = \int a^{-1}dt$, and
\bea
&& z_{\R}^2 = \frac{a^2(\o\dot{\phi}^2 -
12I\dot{\xi}H^2)}{(H + I)^2}, \\
&& c_{\R}^2 = 1 + \frac{-16I\dot{\xi}\dot{H} +
8I^2(\ddot{\xi}-\dot{\xi}H)}{\o\dot{\phi}^2-12I\dot{\xi}H^2},
\label{cr2}
\ena
with $I=-2\dot{\xi}H^2/(1-4\dot{\xi}H)$. 

If we can write $z_{\R} = Q_{\R} |\tau|^{1/2 - \nu_{\R}}$, with $Q_{\R}$ and $\nu_{\R}$
constant, we find $z''_{\R}/z_{\R} = (\nu_{R}^2-1/4)/\tau^2$.
Additionally, if $c_{\R}^2$ is a positive constant, the general solution of Eq.~(\ref{spm})
is a linear combination of Hankel functions
\bea
v = \frac{\sqrt{\pi |\tau|}}{2} e^{i(1+2\nu_{\R})\pi/4}
 \left[c_1 H_{\nu_{\R}}^{(1)}(c_{\R} k|\tau|)
 + c_2 H_{\nu_{\R}}^{(2)}(c_{\R} k|\tau|)\right].
\ena
We choose $c_1=0$ and $c_2=1$, so that positive frequency solutions
in the Minkowski vacuum are recovered in an asymptotic past.
Since $H_{\nu_{\R}}^{(2)}(c_{\R}k|\tau|) \to (i/\pi)\Gamma(\nu_{\R})
(c_{\R}k|\tau|/2)^{-\nu_{\R}}$ for long wavelength perturbations
($c_{\R}k|\tau| \ll 1$), the curvature perturbation after crossing of the Hubble
radius is given by 
\bea
{\R} = \frac{v}{z_{\R}} = e^{i(3+2\nu_{\R})\pi/4} \frac{
 c_{\R}^{-\nu_{\R}}}{4Q_{\R}}
 \frac{\Gamma(\nu_{\R})}{\Gamma(3/2)} \left(\frac{k}{2}\right)^{-\nu_{\R}}.
\ena
The power spectrum of curvature perturbations, ${\cal P}_{\R}=k^3|{\R}|^2/2\pi^2$, becomes
\bea
{\cal P}_{\R} =
\frac{c_{\R}^{-2\nu_{\R}}}{4 \pi^2 Q_{\R}^2}
\left(\frac{\Gamma(\nu_{\R})}{\Gamma(3/2)}\right)^2
\left(\frac{k}{2}\right)^{3-2\nu_{\R}}, 
\ena
with spectral index 
\bea
\label{spi}
n_{\R}-1=3-2\nu_{\R}.
\ena

The Fourier modes of tensor perturbations satisfy~\cite{car01}
\bea
u'' + \left(c_{T}^2 k^2 - \frac{z''_{T}}{z_{T}}\right)u = 0,
\ena
where
\bea
&& z_{T}^2 = a^2(1 - 4\dot{\xi}H) , \\
&& c_{T}^2 = 1 - \frac{4(\ddot{\xi}-\dot{\xi}H)}{1 - 4\dot{\xi}H}.
\label{ct2}
\ena
As above, if $c_{T}^2$ is a positive constant and if $z_T$ can be written as 
$z_{T} = Q_T |\tau|^{1/2 - \nu_T}$ with constant 
$Q_T$ and $\nu_T$, the power spectrum of tensor
perturbations, ${\cal P}_{T}=2k^3|2u/z_{T}|^2/2\pi^2$, is given by
\bea
{\cal P}_T = \frac{8c_T^{-2\nu_T}}{4\pi^2 Q_T^2}
\left(\frac{\Gamma(\nu_{T})}{\Gamma(3/2)}\right)^2
\left(\frac{k}{2}\right)^{3-2\nu_T}.
\ena
The spectral index of the gravitational wave power spectrum is
\bea
n_T=3-2\nu_T.
\label{spit}
\ena
An important observational quantity is the tensor-to-scalar ratio,
\bea
r \equiv \frac{{\cal P}_{T}}{{\cal P}_{\R}} = 8 \frac{Q_{\R}^2}{Q_T^2}
\frac{c_{\R}^{2\nu_{\R}}}{c_T^{2\nu_T}} \left(\frac{\Gamma
(\nu_T)}{\Gamma (\nu_{\R})}\right)^2
 \left(\frac{k}{2}\right)^{2\nu_{\R}-2\nu_T}.
\label{ratio}
\ena

\section{Power-law inflation}

In this section we consider power-law solutions given by
\bea
\label{pls}
a \propto t^{1/\c}, \quad H = \frac{1}{\c t}.
\ena
The power-law solution that involves a single power of cosmic time given by~(\ref{pls}) is
\bea
\dot{\xi} = \a t, \quad V = \frac{\b}{\c t^2},
\ena
where $\a$ and $\b$ are constants, which characterize
the contributions from the GB term and the potential, respectively.
In what follows we restrict our attention to a positive scalar potential,
i.e., $\b>0$. The case of
$\a<0$ is permitted as long as the total energy density is positive.
In this case the potential-to-GB ratio, $\lambda=\c^2\b/12\a$,
becomes a constant.
Eliminating the $\o\dot{\phi}^2$ term from Eqs.~(\ref{beq1}) and
(\ref{beq2}), we have
\bea
(\b+1)\c^2 - (2\a+3)\c + 10 \a = 0,
\ena
which has two solutions, $\c_{+}$ and $\c_{-}$,
\bea
\label{root}
\c_{\pm} = \frac{(2\a + 3) \pm \sqrt{(2\a+3)^2-40\a(\b+1)}}{2(\b+1)}.
\ena
An inflationary solution ($\ddot{a} > 0$) is obtained for $0<\c<1$.
The sign of $\o$ depends on the sign of $(\c^2-2\a\c-2\a)$.
In the case $\a \approx 0$ (no GB contribution), 
we choose $\o=1$ and $\c=\c_{+}$, since $\c_{-}=0$. 
The condition for inflation requires $\b>2$.
In the case $\b=0$ (no potential), we choose $\o=-1$ and $\c=\c_{-}$,
since $\c_{+}>1$. 
The condition for inflation requires $\a<1/4$~\cite{guo07}.
Taking the positive sign of $\phi$, we obtain
\bea
&& \phi = \frac{1}{\c}\sqrt{\frac{2|\c^2-2\a\c-2\a|}{\c}} \ln t, \\
&& \xi(\phi) = \frac{\a}{2} \exp
\left(\sqrt{\frac{2\c}{|\c^2-2\a\c-2\a|}}\,\c\phi\right), \\
&& V(\phi) = \frac{\b}{\c} \exp
\left(-\sqrt{\frac{2\c}{|\c^2-2\a\c-2\a|}}\,\c\phi\right),
\ena
which indicate that an exponential potential and an exponential coupling
give rise to exact power-law solutions. 

\begin{figure}
\includegraphics[height=7cm]{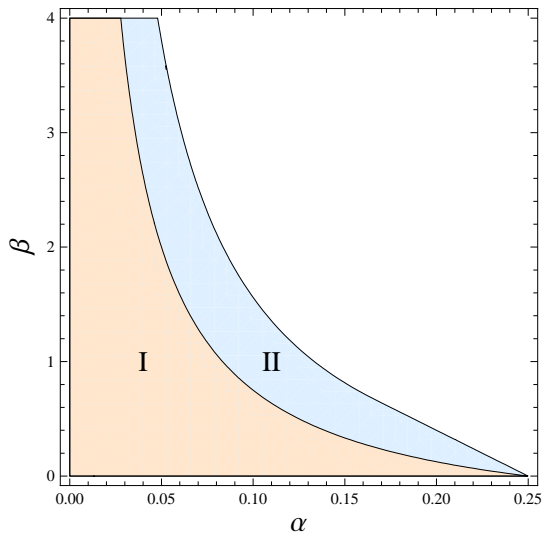}
\caption{Parameter space $(\a,\b)$ for inflationary solutions
with $0<\c_{-}<1$. 
In region I, tensor modes are unstable on small scales, while scalar modes are
stable. The opposite holds in region II.}
\label{gm}
\end{figure}

Let us derive the spectral indices of scalar and tensor
perturbations.
From Eqs.~(\ref{cr2}) and (\ref{spi}) for the scalar perturbations we find
\bea
&& c_{\R}^2 = 1 + \frac{16\a^2(-\c^2 + 5\a\c -
\a)}{\left[(\c^2 - 2\a\c - 2\a)(\c-4\a) + 12\a^2\right](\c-4\a)}, \\
&& n_{\R}-1 = 3 - \left|\frac{3-\c}{1-\c}\right|,
\label{nr}
\ena
which implies a red spectrum for $0<\c<1$. As the phase speed (speed of sound) 
is constant for power-law inflation, the result for the power spectrum obtained above applies.

From Eqs.~(\ref{ct2}) and (\ref{spit}) we find for the tensor perturbations
\bea
&& c_T^2 = 1 - \frac{4\a(\c-1)}{\c-4\a}, \\
&& n_T = 3 - \left|\frac{3-\c}{1-\c}\right|, 
\label{nt}
\ena
the same spectral index as for the scalar perturbations.
The tensor-to-scalar ratio (\ref{ratio}) reads
\bea
r = 16 \frac{(\c^2-2\a\c-2\a)(\c-4\a) + 12\a^2}{(\c-6\a)^2} \left(\frac{c_{\R}^2}{c_{T}^2}\right)^{\nu_{\R}}.
\label{ratios}
\ena
Note that the tensor-to-scalar ratio is independent of $k$, as in minimally coupled power-law models.
However, the phase speed of scalar and tensor perturbations is different from the speed of light in general, e.g.~for the case $1/4 > \a \gg \c \approx 0$ we find $c_R \approx \sqrt{6/5}$ and 
$c_T \approx 0$. 

\begin{figure}
\includegraphics[height=7cm]{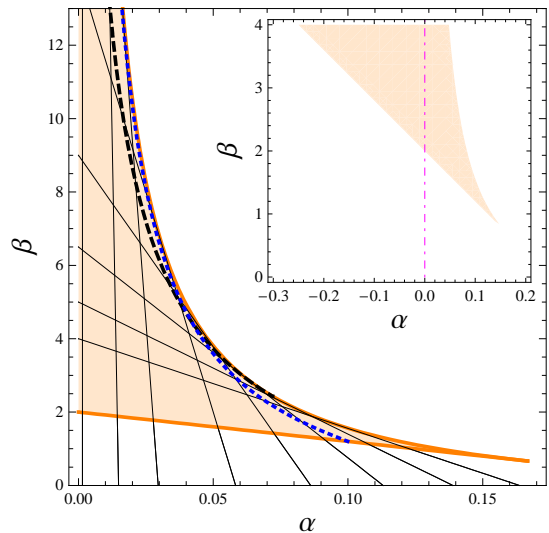}
\caption{Parameter space $(\a,\b)$ for inflationary solutions with
$0<\c_{+}<1$.
In the shaded region both scalar and tensor modes are stable on small scales.
Solid lines correspond to
$\c=0.005$, 0.05, 0.1, 0.2, 0.3, 0.4, 0.5, 0.6
(from bottom left to right).
Models with $\o=-1$ ($\o=1$) are above (below) the black dashed line.
The blue dotted line represents the potential-GB equality ($\lambda=1$).
The potential-dominated (GB-dominated) models are below (above) this line.}
\label{gp}
\end{figure}

\begin{figure}
\includegraphics[height=7cm]{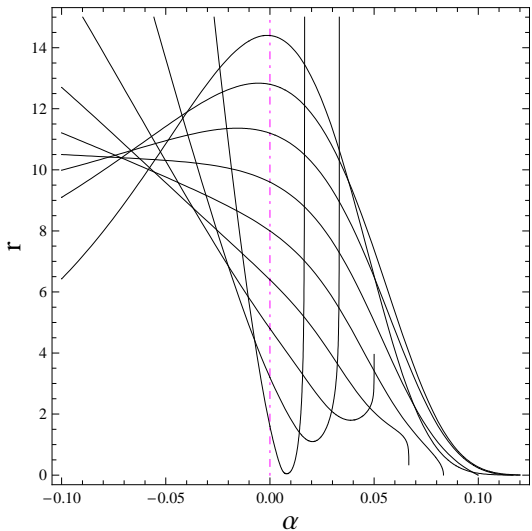}
\caption{Tensor-to-scalar ratio $r$ versus GB-coupling parameter $\a$ for 
inflationary solutions with $\c=0.1$, 0.2, 0.3, 0.4, 0.5, 0.6, 0.7, 0.8,
0.9 (from bottom to top along the dash-dotted line).}
\label{ra}
\end{figure}

If $\a=0$, we have $c_{\R}^2=c_{T}^2=1$. If on top $0<\c\ll1$,
$n_{\R}-1=n_{T} \approx -2\c$, which yield nearly scale-invariant
scalar and tensor perturbations, and $r=-8n_{T}$ --- the standard
consistency relation~\cite{bas05} for single-field slow-roll inflation.
If $\b=0$ and $0<\c<1$, it was shown that the tensor
perturbations suffer from instabilities on small scales, because
$c_{T}^2=2\c-5$ is negative~\cite{guo07}. This type of instabilities
has been also found in Refs.~\cite{cal06}.

In Figure~\ref{gm} we show the parameter space $(\a,\b)$ for inflationary
solutions with $0<\c_{-}<1$. In region I, scalar modes are stable
while tensor modes are unstable on small scales ($c_{\R}^2>0$, but 
$c_T^2<0$). In region II, tensor modes are stable, while scalar modes
are unstable on small scales, since now $c_T^2>0$ and $c_{\R}^2 < 0$.
Unless the system is initially in a Minkowski vacuum state
(in which both $c_{\R}^2$ and $c_T^2$ are positive) before the GB
correction becomes important, a problem arises when we quantize
scalar and tensor modes. We conclude that all solutions involving 
$\gamma_-$ are excluded by stability arguments.

\begin{figure}
\includegraphics[height=7cm]{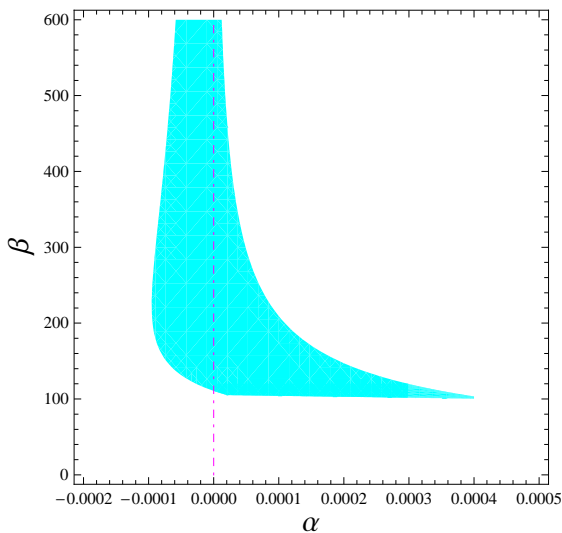}
\caption{Observational constraints on the inflation-model parameters
from the WMAP 5-year analysis. In the shaded region,
$0.942<n_{\R}<1$ and $0<r<0.43$.}
\label{wmapr}
\end{figure}

In Figure~\ref{gp} we show the parameter space $(\a,\b)$ for inflationary
solutions with $0<\c_{+}<1$. Now, both
$c_{\R}^2$ and $c_T^2$ are positive.
We show the potential-dominated (GB-dominated) region
with $\o=1$ ($\o=-1$).
Most of the shaded region corresponds to the potential-dominated
inflationary solution with $\o=1$.

As shown in Figure~\ref{ra}, for inflationary solutions with
$\c>0.5$, there is a small range with $\a>0.08$ in which the
tensor-to-scalar ratio is compatible with observations. However, in that case 
the power spectra of curvature and tensor perturbations are far from
scale-invariant. In order to obtain nearly scale-invariant
spectra of density perturbations, $\c \ll 1$ is required from
(\ref{nr}) and (\ref{nt}). This can be realized when the potential
term is dominant (see also Figure~\ref{gp}). As we see from Fig.~\ref{ra}, 
at small values of $|\a|$, the GB correction leads to a reduction of the 
tensor-to-scalar ratio if $\a>0$, while it is enhanced for $\a<0$.

The recent publication of data from the Wilkinson Microwave
Anisotropy Probe (WMAP)~\cite{kom08} has brought the
global cosmological dataset to a precision where it seriously
constrains inflationary models.
We place constraints on the model parameters by using the results of
the WMAP team.
Fitting the concordance model with power-law spectra for scalar and tensor perturbations 
to WMAP 5-year data alone, yields $n_{\R}=0.986\pm0.022$ (68\% CL)
and $r<0.43$ (95\% CL).
Figure~\ref{wmapr} shows the observational constraints  from WMAP 5-year data on the
parameters ($\alpha$ and $\beta$) of power-law inflation with GB  term.
In the shaded region, $0.942<n_{\R}<1$ and $0<r<0.43$ are 
assumed.

We see that WMAP data allow us to obtain rather tight constraints on the magnitude of 
a possible non-minimal coupling of the inflaton to the GB term ($-1
\times 10^{-4} \lesssim \a \lesssim 4 \times  10^{-4}$ from WMAP, 
while $\a < 1/4$ is possible in principle) because the constraint,
$0.942 < n_{\R} < 1$, requires that $\c<0.028$ from~Eq.(\ref{nr}).
In this range the tensor-to-scalar ratio becomes sensitive to the
parameter $\a$ as shown in Figure~\ref{ra}.

\section{Conclusions}

We have studied inflationary solutions with a non-minimally coupled Gauss-Bonnet term. 
We find that power-law solutions,  giving rise to an exponential potential and an exponential
GB coupling, are in agreement with observation.
The important quantities, directly linked to observations, are
the spectral indices $n_{\R}$ and $n_{T}$, together with the
tensor-to-scalar ratio $r$. Provided that $c_{\R}^2$ and $c_{T}^2$ are
positive constants in the mode equations, we obtain the spectral
indices of scalar and tensor perturbations given by Eqs.~(\ref{nr}) and
(\ref{nt}), respectively, and the tensor-to-scalar ratio~(\ref{ratios}).

Given the model parameter $\a$ and $\b$, there are in general two
solutions, $\c_{-}$ and $\c_{+}$. 
Although scale-invariant spectra are generated for density
perturbations, either scalar or tensor perturbations are faced with
ultraviolet instabilities associated with negative
$c_{\R}^2$ or $c_{T}^2$ in the shaded region
of Figure~\ref{gm}. In the shaded region of Figure~\ref{gp},
this type of instabilities can be avoided. When the potential dominates,
it is possible to generate
nearly scale-invariant curvature perturbations with a reduced or
enhanced tensor-to-scalar ratio in the case
of $\a>0$ or $\a<0$, respectively.

In this work we restricted our attention to power-law solutions, which is intended to be a first 
step towards a more complete understanding of the influence of a non-minimal coupling to the 
Gauss-Bonnett term. The comparison to WMAP data showed that the cosmic microwave sky 
provides us with a mean to strongly constrain the magnitude of that coupling. The next step will 
be the investigation of the slow-roll regime for an arbitrary form of the coupling and an arbitrary 
potential.

\begin{acknowledgments}
We thank Christophe Ringeval, Shinji Tsujikawa and David Wands
for valuable discussions.
This work was supported by the Alexander von Humboldt Foundation.
\end{acknowledgments}

\end{document}